\documentclass[journal=jacsat,manuscript=article]{achemso}
\usepackage[utf8]{inputenc}
\setkeys{acs}{articletitle = true}
\setkeys{acs}{maxauthors=10,etalmode=truncate}
\usepackage{chemformula} % Formula subscripts using \ch{}
\usepackage[T1]{fontenc} % Use modern font encodings
\usepackage{enumitem}
\usepackage{graphicx}
\usepackage{amsmath}
\usepackage{amssymb}
\usepackage{bm}
\usepackage{titlecaps}
\Addlcwords{a an the of on at with for and in into from by without}
\SectionNumbersOn

\usepackage{rotating}

\author{Markus
  Meuwly}\email{m.meuwly@unibas.ch}\affiliation[University of Basel]
       {Department of Chemistry, University of Basel,
         Klingelbergstrasse 80, 4056 Basel, Switzerland}
       \alsoaffiliation[Brown University]{Department of Chemistry,
         Brown University, Providence/RI, USA}

\title[]{Transformative Applications of Machine Learning for Chemical
  Reactions}

\begin{document}

\begin{abstract}
Machine learning techniques applied to chemical reactions has a long
history. The present contribution discusses applications ranging from
small molecule reaction dynamics to platforms for reaction
planning. ML-based techniques can be of particular interest for
problems which involve both, computation and experiments. For one,
Bayesian inference is a powerful approach to include knowledge from
experiment in improving computational models. ML-based methods can
also be used to handle problems that are formally intractable using
conventional approaches, such as exhaustive characterization of
state-to-state information in reactive collisions. Finally, the
explicit simulation of reactive networks as they occur in combustion
has become possible using machine-learned neural network
potentials. This review provides an overview of the questions that can
and have been addressed using machine learning techniques and an
outlook discusses challenges in this diverse and stimulating field. It
is concluded that ML applied to chemistry problems as practiced and
conceived today has the potential to transform the way with which the
field approaches problems involving chemical reactions, both, in
research and academic teaching.
\end{abstract}

\maketitle

\section{Introduction}
The prediction of chemical reaction outcomes in terms of products,
yields, or reaction rates from computations is a formidable
undertaking. Characterizing and understanding chemical transformations
is at the heart of chemistry and provides the necessary information
about mechanisms and efficiencies of such processes. A quantitative
understanding of the speed and efficiency of chemical reactions
pertains to, but is not limited to, fields as diverse as
atmospheric\cite{vereecken2018perspective},
combustion\cite{klippenstein:2017}, or astrophysical
reactions\cite{wakelam:2012} to explosions and
enzymatic\cite{mulholland:2013} or organic
reactions.\cite{cheong:2011} It is both of interest to unravel
mechanistic underpinnings of known and well-characterized reactions
and to predict reaction outcomes based on physically based or
physics-inspired models\cite{MM.nonconv:2020} or on high-accuracy
PESs.\cite{unke:2020,manzhos:2020,guo.3:2020}\\

\noindent
If one is interested in following bond breaking and bond formation in
time and space, there are in general two possibilities based on
dynamics simulations: those directly and explicitly using quantum
mechanical (QM) calculations for the total (electronic) energy and
methods based on representations of these energies such as
(parametrized) empirical force fields or - as done more recently -
based on machine learning. Empirical energy functions represent the
potential energy surface (PES) as a computable function for given
atomic coordinates whereas QM-based methods solve the electronic
Schr\"odinger equation directly for every configuration $\vec{x}$ of
the system for which it is required. QM-based
methods\cite{qiang.jcp:2016} provide the most general framework for
investigating the dynamics of chemical reactivity without preconceived
molecular structures or reaction mechanisms. However, there are
certain limitations which are due to the computational approach {\it
  per se} (such as the speed and efficiency of the method) or due to
practical aspects of quantum chemistry (e.g. basis set superposition
error, the convergence of the Hartree-Fock wavefunction to the desired
electronic state for arbitrary geometries, or the choice of a suitable
active space irrespective of molecular geometry). Improvements and
future avenues for making QM-based approaches even more broadly
applicable have been recently discussed.\cite{qiang.jcp:2016}\\

\noindent
For larger systems, such as reactions involving biomolecules, mixed
quantum mechanics/molecular mechanics (QM/MM) treatments have become
popular.\cite{senn:2009} In this approach the system is decomposed
into a ``reactive region'' which is treated with a quantum chemical
(or semiempirical) method whereas the environment is described by an
empirical force field. This makes the investigation of certain
processes feasible that would not be amenable to a full quantum
treatment such that even free energy simulations in multiple
dimensions can be carried out.\cite{cui:2016} One of the current open
questions concerns the size of the QM region required for converged
results which was recently considered for Catechol
O-Methyltransferase.\cite{kulik:2016}\\

\noindent
The use of empirical force fields to follow chemical reactions dates
back at least 50 years.\cite{ellison:1963,ellison.2:1963,kar65:3259}
Such an approach has seen various incarnations, including the theory
of diatomics in molecules\cite{ellison:1963,ellison.2:1963}, empirical
valence bond (EVB)\cite{warshel:1980} with its extension to several
bonding patterns specifically for proton transport in
water,\cite{voth:1998} the ReaxFF force field\cite{goddard:2001}, the
reactive molecular dynamics force field (RMDff) initially developed
for polymers and based on the concept of bond
order,\cite{forney:2003,westmoreland:2007}, or adiabatic reactive
molecular dynamics (ARMD)\cite{MM.armd:2006,MM.armd:2008} together
with its multi-state
variants.\cite{MM.armd:2014,MM.armd:2018,MM.pt:2019} There is a broad
range of reviews on the subject of investigating chemical reactions
based on established treatments\cite{farah:2012} of the potential
energy surfaces with applications in gas
phase,\cite{collins:2002,bowman:2011}
solution,\cite{hynes:1985,hynes:2015} and enzymatic
reaction\cite{warshel:2003,truhlar:2006,himo:2017,mulholland.2:2018,mulholland:2020}
dynamics and the technology has also been extended to coarse-grained
simulations.\cite{bohm:2011}\\

\noindent
The present work focuses on more recent developments and applications
of machine learning techniques applied to problems involving reactions
in the gas phase, in solution, and in enzymes. Most problems
concerning the {\it representation} of the underlying potential energy
surfaces are excluded, as these are already well covered by
accompanying contributions to this special
issue.\cite{unke:2020,manzhos:2020} Rather, the present work focuses
on the application of ML-based techniques to problems that can not
otherwise be exhaustively sampled, on the interplay between
experimental observables and computations and how to exploit this for
improved understanding of intermolecular interactions, and on dealing
with entire reaction networks.\\

\noindent
Machine Learning (ML) is a data-driven method based on statistical
learning theory to generate numerical models that generalize to new
data, not used in the learning process.\cite{vapnik1998} Ideally, ML
models interpolate and extrapolate to new data but in general this
needs to be verified for every task. Historically, ML can be traced
back to work on ``Turing's Learning Machine''. In 1951 the
``Stochastic Neural Analog Reinforcement Calculator'' (SNARC) was
built by Minsky and Edmonds as a summer research project which is
considered one of the first ``artificial neural network''
(ANN). Limitations of the learning capabilities of ANNs were described
in ``Perceptron''\cite{minsky:1969} and convolutional
(CNN)\cite{fukushima:1980} and recurrent (RNN)\cite{hopfield:1982} NNs
were developed subsequently. In the 1970s automatic differentiation
was developed\cite{linnainmaa:1976} which eventually lead to
backpropagation that allows to learn internal
representations\cite{williams:1986} and was first applied to NNs in
the behavioural sciences.\cite{werbos1975} With the possibility to
compute extensive reference data sets, application of ML-based
techniques to questions of chemical reactivity has become a powerful
complement to established methods. \\

\noindent
One aspect of particular interest in the present work concerns the
interplay between observation and information obtained from a computer
model. ``Observation'' in the present context can either be an
experimental observation or one from another computation. As an
example, the PES can be a parametrized or non-parametric
representation of {\it ab initio} computed energies and the
observables can be obtained from either a classical, quasiclassical or
quantum nuclear dynamics simulation such as an inelastic scattering
cross section. Conversely, it is also possible to start with a set of
experimental observations, for example reaction rates, and aim at
determining the corresponding underlying PES that supports these
experimental observables.\\

\section{Machine Learning of Reaction Observables}
Machine Learning techniques can be used to develop comprehensive
models for observables that are directly related to
experiments. Examples include prediction of
thermal\cite{bowman.reaction:2019,bowman.reaction:2020,komp:2020} or
quantum reaction rates\cite{komp:2020}, state-to-state cross
sections\cite{MM.cross:2019}, mapping initial to final state
distributions\cite{MM.cross:2020}, or even chemical reaction
yields.\cite{schwaller:2020}\\

\subsection{Learning Reaction Rates}
Chemical reactions involve bond-breaking and bond-formation
processes. Hence, if the explicit nuclear dynamics describing the
transition between reactant and one or several possible products is of
interest, the motion between the two involves transgressing a barrier
on the multidimensional PES. Empirical force fields, originally based
on experimental information such as structure, spectroscopy and
thermodynamics\cite{oplsff,charmm22,amberff,gromos:2004} are not
suitable for following chemical reactions as the bonding pattern
between the atoms is fixed. With the advent of efficient electronic
structure methods, attention has shifted to either complement existing
parametrizations with information from (high-level) electronic
structure
data\cite{ponder:2010,MM.mtp:2012,ren:2013,MM.mtp:2013,headgordon:2020,MM.nonconv:2020}
or to develop models that are entirely based on quantum chemical
calculations.\cite{schmidt:2014}\\

\begin{figure}
\includegraphics[scale=0.67]{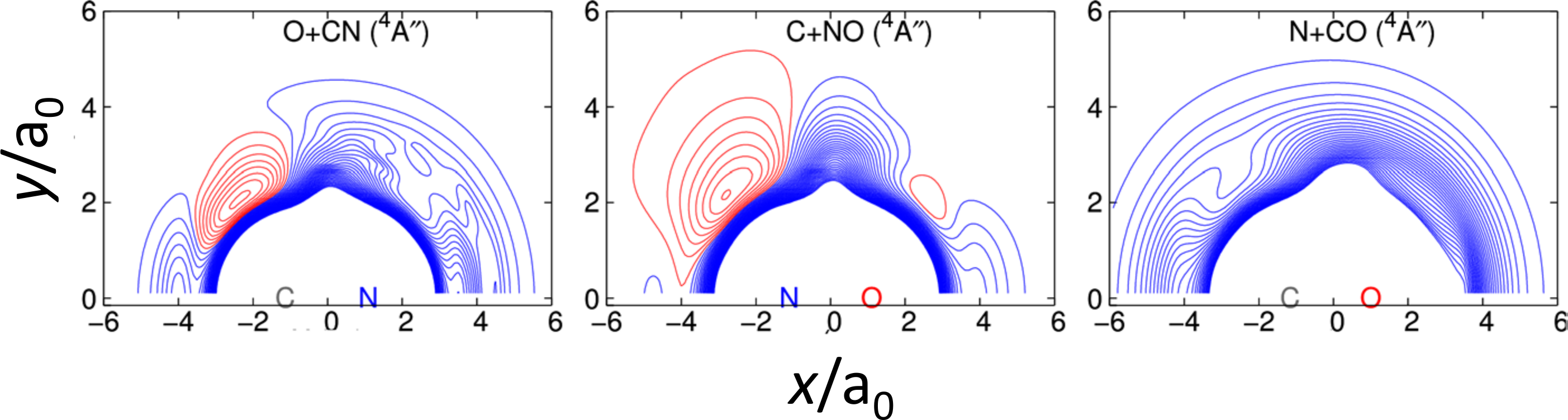}
\caption{Contour plots of the $^4$A'' PESs for the [CNO] system
  represented as a reproducing kernel Hilbert
  space.\cite{hollebeek:1999,unk17:1923} The diatoms (CN, NO, CO from
  left to right) are at their equilibrium structures (2.234, 2.192 and
  2.150 a$_0$, respectively). The spacing between the isocontours is
  0.2 eV with red lines corresponding to negative energies (--0.1,
  --0.3, --0.5 eV and lower) and blue lines to positive energies (0.1,
  0.3, 0.5 eV and higher).}
\label{fig:cno.pes}
\end{figure}

\noindent
From a microscopic perspective suitable methods to follow chemical
transformations - e.g. quasi classical trajectory or quantum
simulations - at a molecular length scale are sensitive to the {\it
  entire} PES and require a global, reactive PES. Representing such a
PES can be very challenging even for triatomic
systems\cite{guo:2017,MM.reactions:2020} because their topographies
can be rather complex.\cite{MM.cno:2018} An example for such a PES for
the [CNO] reactive system is reported in Figure
\ref{fig:cno.pes}. These 3-dimensional PESs were represented as a
reproducing kernel Hilbert space which exactly matches the reference
{\it ab initio} calculations on the reference
points.\cite{hollebeek:1999,unk17:1923} Finding parametrized functions
with similar performance is in general very challenging. An
alternative are permutationally invariant
polynomials.\cite{qu18:151}\\

\noindent
The determination of accurate reaction rates from computations is a
formidable task even for seemingly simple A+BC$\rightarrow$AB+C atom
exchange reactions\cite{MM.reactions:2020} such as the
O+CN$\rightarrow$[C+NO,N+CO] reaction for which PESs are illustrated
in Figure \ref{fig:cno.pes}. Modern approaches for gas phase reactions
are based on a) the computation of thousands of energies at a high
level of quantum chemical theory (e.g. coupled cluster (CCSD(T)) or
multi reference (MRCI) treatments) using large basis sets, b) the
representation of these energies either as a parametrized
function\cite{qu18:151} or using machine learning techniques such as
neural
networks\cite{beh07:146401,schutt:2018,unk19:3678,mm.ht:2020,unke:2020},
reproducing kernel Hilbert space
methods\cite{ho96:2584,hol97:7223,hollebeek:1999,unk17:1923}, and c)
following the nuclear dynamics either using classical mechanics
(quasiclassical trajectory calculations (QCT))\cite{kar65:3259,tru79}
or solving the three-dimensional nuclear Schr\"odinger equation. Such
an approach is overall very computationally demanding and the quality
of the computed rates depends sensitively on the accuracy of the
underlying PES.\cite{panesi:2020}\\

\noindent
In an effort to alleviate this problem, a recent
effort\cite{bowman.reaction:2019} explored the possibility to use
Gaussian Process (GP) regression to train a correction $\chi(T)$ to
predict thermal rates $k(T)$ in
\begin{equation}
  k(T) = [\kappa_{\rm ECK}(T) k^{\rm TST}(T)] \chi(T)
\label{eq:ml.reaction}
\end{equation}
Here, $k(T)$ was determined for $\sim 50$ different reactions based
primarily on collinear collisions - which is a simplification but does
not limit the general applicability of the approach - and the rates
$\kappa_{\rm ECK}(T)$ and $k^{\rm TST}(T)$ are those from a simplified
treatment of the Eckart tunneling correction (ECK) to conventional
transition state theory (TST). Training was done for 13 reactions and
the model was tested on 40 reactions. Between 3 and 5 descriptors were
chosen to represent $\chi(T)$ and it was reported that judicious
choice of the descriptors can lead to marked improvements in the model
performance.\\

\noindent
Compared with either TST or the Eckart treatment, the machine learned
model performed best on the training set, on systems with symmetrical
and asymmetrical barriers. In all cases, the error of the learned
model ranged from 10 \% to 120 \% when compared with the exact
data. This is a marked improvement over errors ranging from 80 \% to
180 \% (for ECK) and 180 \% to 760 \% (for TST). For reactions
involving hydrogen abstraction from CH$_4$ the Eckart model was best,
followed by the GP-trained model and TST. This study also highlights a
critical need for both, the quality and quantity of reliable reference
data in training such models. One surprising finding is the fact that
using a model trained on collinear reactions (``2d'') but applying is
to a data set for reactions in full dimensionality (``3d'') can
perform quite well.\\

\noindent
A recent application of this model concerned the
O($^3$P)+HCl$\rightarrow$OH+Cl reaction\cite{bowman.reaction:2020}
which can be considered as a particularly challenging example due to
the large reaction barrier, the presence of low-energy reactive
resonances and the heavy-light-heavy character of the system. Future
applications of such an approach may be possible on sufficiently
extensive and curated data sets from actual experiments. As the
reference data originated from a broad range of PESs it would also be
of interest to determine whether model prediction can be further
improved if all PESs are based on the same or a similar level of
theory.\\

\noindent
Machine learning approaches were also extended to one-dimensional
quantum reaction rates.\cite{komp:2020} Based on a large number of
quantum rates $k_{\rm Q}(T)$, a deep neural network (DNN) was
trained. The potential energy surfaces considered include single and
double barriers, symmetric and asymmetric shapes. The optimized DNN,
trained on $\sim 1.5$ million data points, was then applied to
predicted rates for a range of gas phase and surface reactions. The
overall accuracy on the test set for $\log{k_{\rm Q}(T)}$ was 1.1 \%
and even at temperatures below 300 K for which tunneling effects ae
expected to become important, the relative error was only $\sim 30$
\%.\\

\subsection{State-to-State Models and Rates}
Maintaining the full dimensionality of the problem, a NN-based,
machine learned model was developed for the state-to-state (STS) cross
sections of the N($^4$S)+NO($^2\Pi$) $\rightarrow$
O($^3$P)+N$_2$(X$^1\Sigma_g^+$) reaction.\cite{MM.cross:2019} These
reactions are relevant in the atmosphere and for hypersonic
flight.\cite{dsmc:2017} For the $^3$A' state the total state
space for this problem, i.e. the total number of state-to-state
transitions, involves a maximum of 47 and 57 vibrational states for NO
and N$_2$, respectively, and maximum rotational quantum numbers for NO
and N$_2$ of 241 and 273. Hence, there are 6329 ro-vibrational states
for the N+NO channel, and 8733 states for the O+N$_2$ channel. To
converge one specific state-to-state cross section $\sigma_{v,j
  \rightarrow v'j'} (E_t)$ for given translational energy $E_t$
typically $10^4$ to $10^5$ QCT simulations need to be run. Doing this
for the $10^4$ initial states going into all $10^4$ final states would
require an estimated $10^{12}$ to $10^{13}$ QCT trajectories for
directly sampling all available ro-vibrational states which is
computationally impractical, even for this low-dimensional 3-body
system. For diatom-diatom collisions the number of transitions
increases to $\sim 10^{15}$ and the necessary number of QCT
simulations approaches $10^{20}$.\cite{grover:2019} In such
situations, ML approaches can provide an alternative to address the
problem of characterizing product distributions from given reactant
state distributions or more generally to determine final distributions
from initial distributions given specific evolution equations or
rules. In previous work\cite{MM.cross:2019}, a model for
state-to-state (STS) cross sections from quasi classical trajectory
(QCT) simulations of an atom-diatom collision system using a neural
network (NN) has been proposed.  \\

\noindent
{\bf State-to-State Model} The NN architecture for this application
was based on ResNet\cite{he16:770} which uses identity shortcut
connections to alleviate the vanishing gradient
problem\cite{glorot2010understanding} that slows down the learning
capacity with increasing depth of the NN. The input is transformed
through four identical residual blocks\cite{he16:770}, after which a
linear transformation followed by a scaled sigmoid function is used to
project onto the final output.\cite{MM.cross:2019} The weight matrices
$\mathbf{W} \in \mathbb{R}^{F \times F} $ (weight matrix) and
$\mathbf{b} \in \mathbb{R}^{F}$ (bias vector) contain the parameters
to be optimized. The residual blocks consist of two dense layers with
the same number of nodes (here $F=24$) and transform their input
$\mathbf{x}^{l}$ according to
\begin{equation}
 {\bf x}^{l+2} = {\bf x}^{l} + {\rm ReLU}[{\bf W}^{l+1}{\rm
     snasinh}({\bf W}^{l}{\bf x}^{l} + {\bf b}^{l}) + {\bf b}^{l+1}],
\end{equation}
The superscript $l$ labels parameters for layer $l$. Two different
activation functions, one for rectified linear units
(ReLU)\cite{nair2010rectified} and a
self-normalizing\cite{klambauer2017self} inverse hyperbolic sine
(snasinh)\cite{unke2018reactive} are used in the residual blocks. The
final output is obtained from
\begin{equation}
y = C \times {\rm sig}(\mathbf{W}^{o}\mathbf{x}^{l} + b^{o}),
\end{equation}
where $C=0.4$ is a scaling constant, $\mathrm{sig}(x)=(1+e^{-x})^{-1}$
denotes the sigmoid function and the superscripts $o$ and $l$ denote
parameters $\mathbf{W}^o \in \mathbb{R}^{F\times 1}$ and $b^o \in
\mathbb{R}$ corresponding to the output layer and the hidden features
$\mathbf{x}^{l}$ obtained after the last residual block,
respectively.\\

\noindent
The 12 input features $\mathbf{f}$ are (i) internal energy, (ii)
vibrational energy, (iii) vibrational quantum number, (iv) rotational
energy, (v) rotational quantum number, (vi) angular momentum of the
diatom, (vii) relative translational energy, (viii) relative velocity,
(ix, x) turning periods of the diatom, (xi) rotational barrier height
and (xii) vibrational time period of the diatom. For state-to-state
cross sections, the same 12 features for the final states of the
products are also included as input (i.e., $N_{\rm in}=24$).  The loss
functions ($L_f$) was
\begin{equation}
L_f = \frac{1}{N}\sum_1^N [{\rm log}(y'+1.0)-{\rm log}(y'+|y-y'|+1.0)]^2,
\label{eq:loss}
\end{equation}
where $y'$ and $y$ are the reference (QCT) and predicted values (NN),
respectively. All parameters of the NN are initialized according to
the Glorot initialization scheme\cite{glorot2010understanding} and
optimized using Adam optimization.\cite{kingma2014adam}\\

\begin{figure}
\includegraphics[scale=0.65]{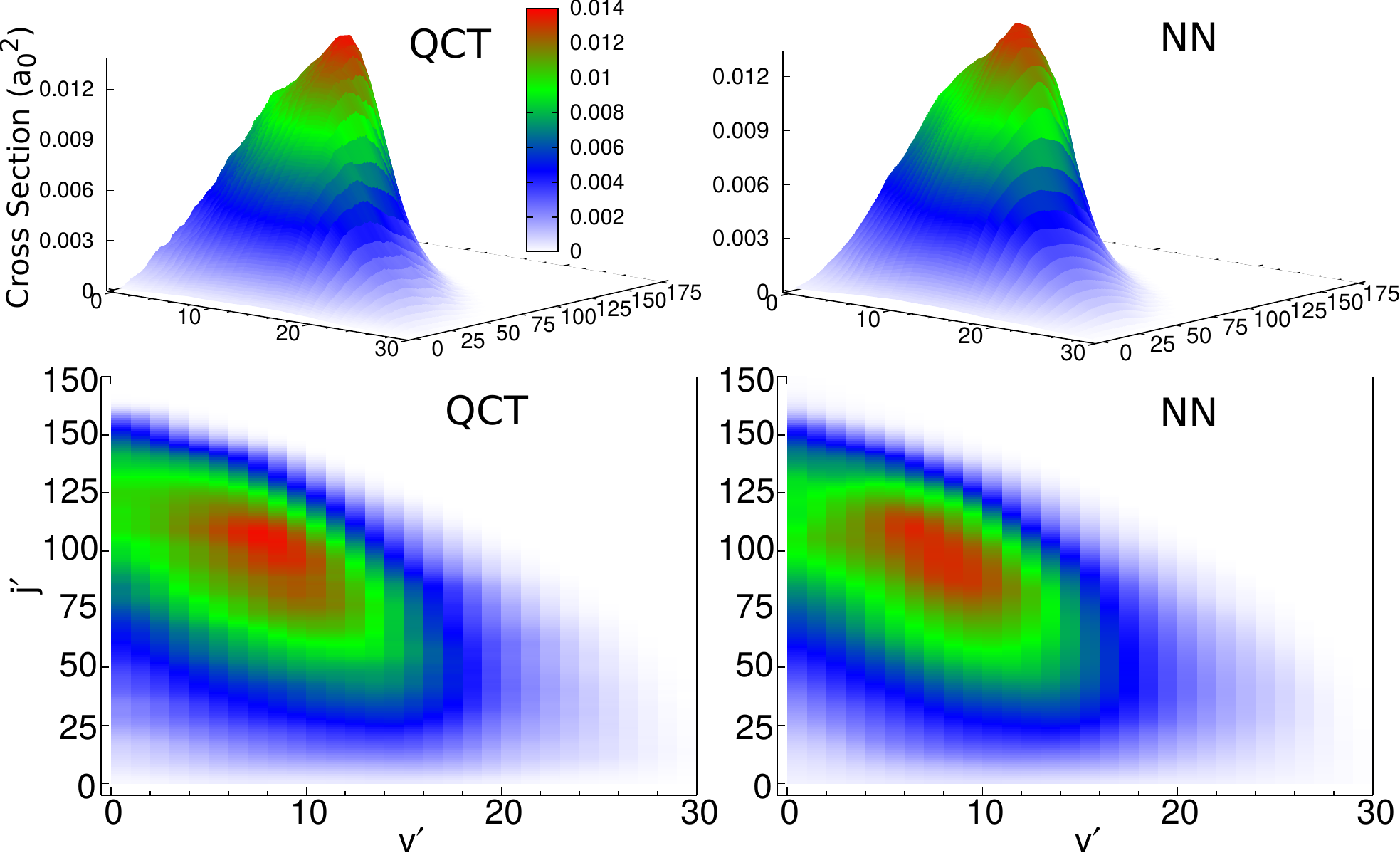}
\caption{3D surface (upper panel) and contour color map (lower panel)
  of QCT calculated and NN-STS predicted state-to-state cross
  sections for N + NO($v=6, j=30$) $\rightarrow$ O + N$_2(v',j')$ at
  $E_t =$ 2.5 eV.}
\label{fig03mm}
\end{figure}

\noindent
To quantify the accuracy of the NN, additional QCT calculations were
performed for independent initial conditions at fixed $E_t$. Total QCT
cross sections are then compared with the NN predictions, see Figure
\ref{fig03mm}. Again, the NN results describe the explicit QCT
simulations which validates the use of such a model to predict
microscopic information for such a reaction.\\

\begin{figure}
\includegraphics[scale=1.0]{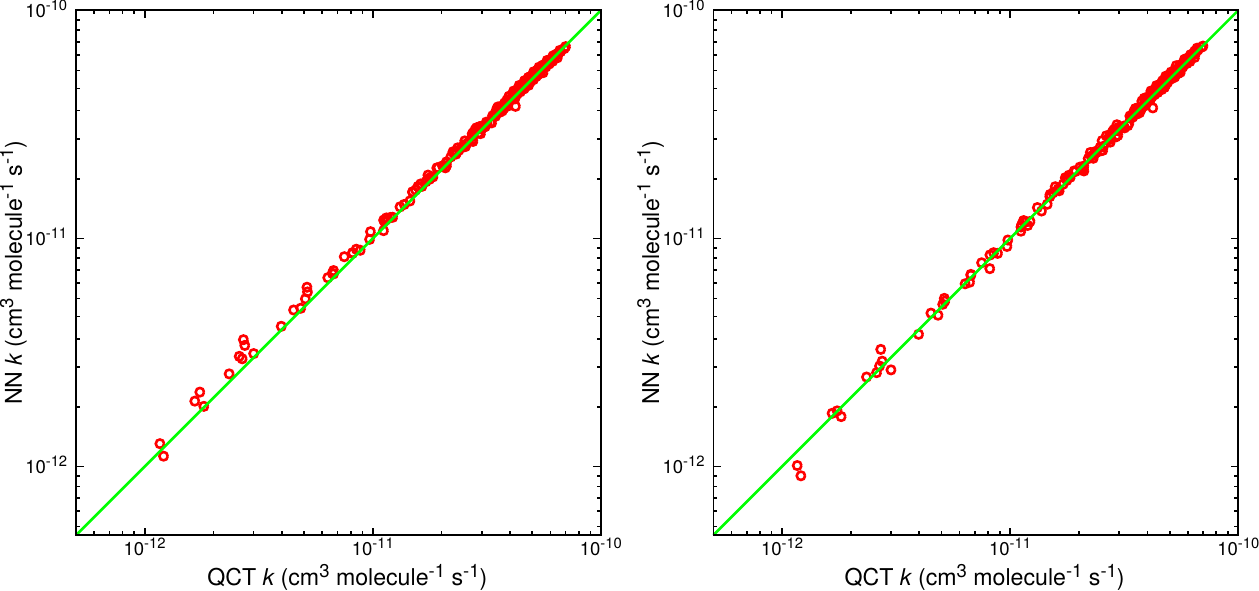}
\caption{Correlation between the QCT-calculated and NN-predicted
  initial state selected rates. Left panel: Initial state selected QCT
  vs NN-STS rates for $v = 5, 10, 15$ and 20, $ j = 20, 25,$ 40, 60,
  85 and 110, and at $T = 2000$, 4000, 6000, 8000, 10000, 12000,
  14000, 16000 and 18000 K. Right panel: Total QCT vs NN-Tot rates at
  $T = 1000$, 2000, 3000, 4000, 5000, 8000, 10000, 12000, 15000, 18000
  and 20000 K. Diagonals are shown as solid lines.}
\label{sfigt11}
\end{figure}

\noindent
Initial state specific rates for the reaction were also calculated at
temperatures between 2000 and 20000 K for a few selected reactant
states using QCT simulations and compared with the rates obtained from
the NN models. The NN models successfully capture the rates from QCT,
see Figure \ref{sfigt11}. Although maximum relative errors of $\sim
17$\%\ are found for specific initial states, in most cases the
relative errors are $< 5$\%. State-specific and total reaction rates
as a function of temperature from the NN are in quantitative agreement
with explicit QCT simulations and confirm earlier simulations and the
final state distributions of the vibrational and rotational energies
agree as well.\\

\noindent
{\bf Distribution-to-Distribution Model} In a recent extension, the
STS model was generalized to a distribution-to-distribution
model\cite{MM.cross:2020} for the relative translational energy
$E_{\rm trans}$, the vibrational $v$ and rotational $j$ states of a
reactive atom-diatom collision system. Hence the NN-based model
trained involved the reactant state distributions
($P(E_{\text{trans}}), P(v), P(j)$ where $(v,j)$ labels the
rovibrational state of the diatom) and predicts the three
corresponding product state distributions ($P(E_{\text{trans}}'),
P(v'), P(j')$) whereby $P(v)$ and $P(j)$ are marginal distributions,
i.e. $P(v)=\sum_{j} P(v,j)$ and $P(j)=\sum_{v} P(v,j)$. Working with
the underlying distributions leads to considerably smaller NNs that
need to be trained which also speeds up the learning process. For this
task, a multilayer perceptron with two hidden layers was used with 10
to 40 input and output nodes depending on the representation of the
distributions. The two hidden layers contain between 6 and 12 nodes
each.\\

\begin{figure}[h!]
\begin{center}
\includegraphics[width=0.99\textwidth]{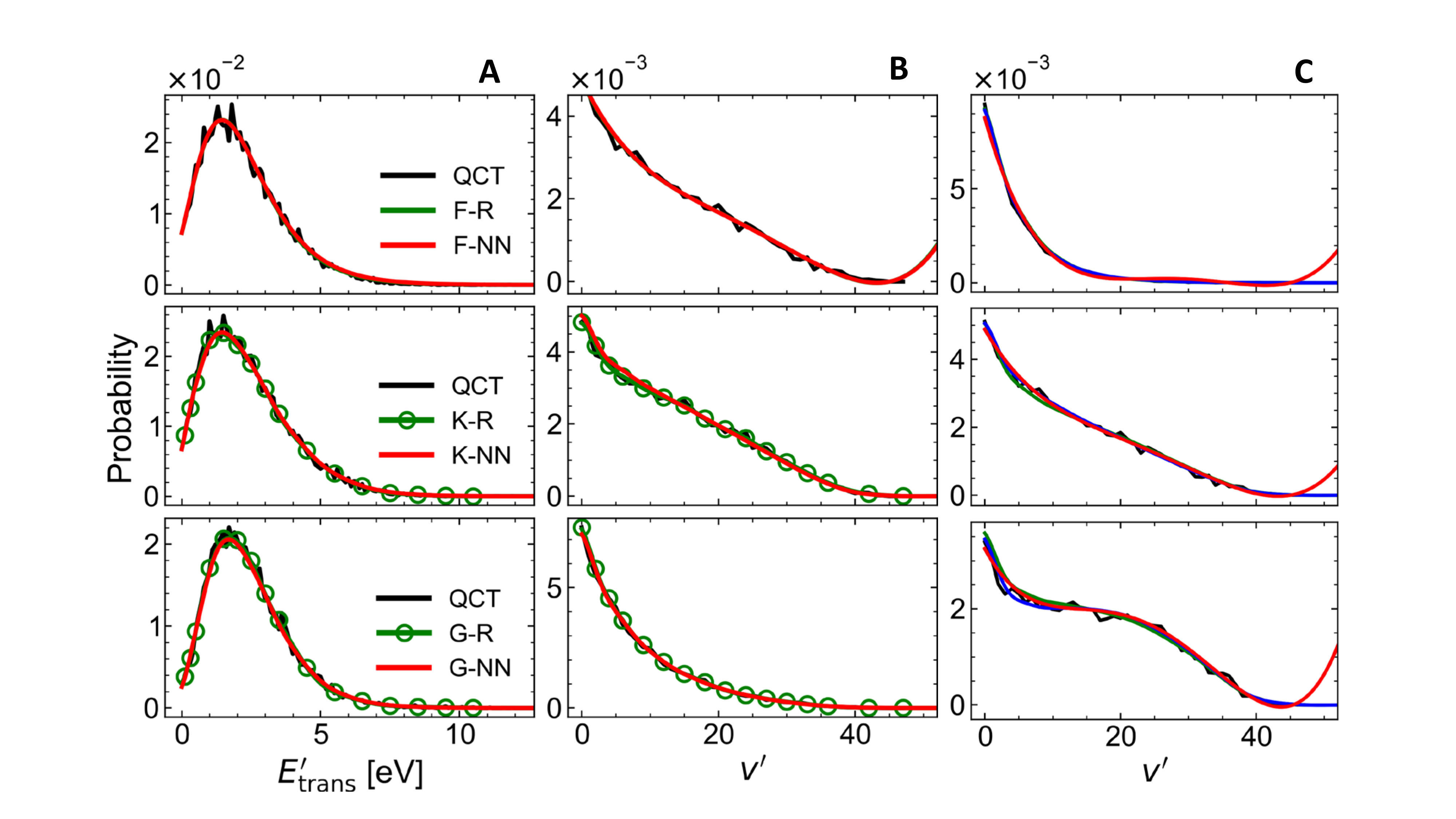}
\caption{Product state distributions from explicit QCT simulations
  (QCT, black traces) as well as the corresponding reference data (-R)
  and the predictions (-NN) from the F-based (top), K-based (middle)
  and G-based (bottom) approaches. Also, the amplitudes to construct
  the reference RKHS-based representations in the K- and G-based
  approaches are reported (circles). For panels A and B the data sets
  are from QCT simulations with initial conditions sampled at
  translational, vibrational, and rotational temperatures $(T_{\rm
    trans}, T_{\rm vib}, and T_{\rm rot})$ for models representative
  of the average performance of the NN for each of the
  representations. These temperatures are (9500 K, 16000 K, 16000 K),
  (10250 K, 19250 K, 19250 K), and (12000 K, 9750 K, 9750 K) and
  [RMSD$_{\text{NN}}=0.0005$, $R^2_{\text{NN}}=0.9995$],
  [RMSD$_{\text{NN}}=0.0013$, $R^2_{\text{NN}}=0.9984$], and
  [RMSD$_{\text{NN}}=0.0009$, $R^2_{\text{NN}}=0.9993$] for F-, K-,
  and G-based representations, respectively. In panel C the final
  vibrational distributions are reported for initial conditions at
  temperatures (12500 K, 5750 K, 5750 K), (9500 K, 16000 K, 16000 K),
  and (5750 K, 19250 K, 19250 K) for the top, middle, and bottom
  panels. Towards the highest values $v'$ the F-based approach (red)
  is unable to correctly model $P(v')$. Note also the stark contrast
  to a typical Boltzmann distribution for the vibrational state
  distribution for the middle and bottom panels in C. Figure adapted
  from Ref.\cite{MM.cross:2020}}
\label{fig:dtd}
\end{center}
\end{figure}

\noindent
For representing the distributions, parametrized functions (F),
ML-representations based on reproducing kernels (K), and the actual
grid points (G) were considered. In general, all three approaches
accurately describe the (equilibrium Boltzmann) reactant state
distributions. However, the product states are nonequilibrium
distributions as they are from high-temperature simulations, ranging
from 2000 K to 20000 K. Figure \ref{fig:dtd} reports final
translational and vibrational distributions $P(E'_{\rm trans})$ and
$P(v')$ from different simulations compared with the machine learned
predictions. For Figures \ref{fig:dtd}A and B the QCT simulations were
from conditions representative of the average $R^2$ over all test data
for the respective method, i.e. using F-, K-, or G-based
representations of the distributions. For conditions most
representative of the average performance, a function-based approach
is somewhat better suited ($R^2 = 0.999$) than a K-based ($R^2 =
0.998$) or a G-based ($R^2 = 0.999$) approach.\\

\noindent
The representation by fitting to parametrized functions (F-based)
leads to differences, in particular for $P(v')$ (e.g., deviations for
small and high $v'$ or extra undulations in Figure
\ref{fig:dtd}C. However, the deviations observed for high $v'$ are
only partially relevant, as the accessible vibrational and rotational
state space is finite, here $v'_{\rm max}=47$, $j'_{\rm
  max}=240$. Considering the K- and G-based approaches, the reference
representations describing product distributions are nearly identical
and reproduce the QCT data very closely, see blue and green traces in
Figure \ref{fig:dtd}C.\\

\subsection{Gaussian Processes and Bayesian Inference}
Gaussian Process (GP) regression is a non-parametric, supervised
learning technique and one of several kernel-based methods to generate
ML models.\cite{gp06} Previously, GP has been applied to regression
and classification problems but more recently it has also been used to
represent intermolecular PESs\cite{krems:2016,guo:2017} and, combined
with Bayesian optimization, to address the inverse problem of reactive
scattering.\cite{krems:2019} The ``inverse problem'' of determining
the interaction potential from scattering data is a long-standing
problem in chemical
physics\cite{firsov:1953,buck:1969,buck:1974,gerber:1979,buck:1986}
and has been handled within Tikhonov regularization and using
minimization procedures for diatomic molecules.\cite{rabitz:1995}\\

\noindent
The problem has been formulated in a combined GP and Bayesian
inference context.\cite{krems.bayes:2019} The formal development
starts from a global, reactive PES expressed as $V(\vec{r}) =
\sum_{i=1}^{N} w_i(\vec{r})E_i$, similar to
MS-ARMD\cite{MM.armd:2014}, where $E_i$ are energies from an {\it ab
  initio} calculation and the weights $w_i(\vec{r})$ are optimized
through GP regression to yield most accurate scattering cross
sections. A GP is entirely specified by the conditional mean
$\mu(\vec{r}_i)$ and the conditional variances $\sigma(\vec{r}_0)$ at
an arbitrary configuration $\vec{r}_0$. The conditional mean and
variances can be represented as an $n-$dimensional vector and a $n
\times n$ square matrix of covariances, respectively.\cite{krems:2019}
The GP model is trained by determining the best covariance matrix
given a set of reference data (e.g. energies from {\it ab initio}
calculations or scattering cross sections). In order to progress, a
model for the covariances needs to be assumed which can be Matern
functions, Gaussians, rational quadratic kernels or other simple
functions.\cite{gp06} The model is then optimized using a suitable
likelihood function which is the log marginal likelihood as is
customary for GP optimization. It is noted that up to this point such
an approach is also reminiscent of the reproducing kernel Hilbert
space technique which employs different classes of functions for
representing the kernel matrix and which has been successfully used to
represent
PESs.\cite{aronszajn1950rkhs,ho96:2584,hol97:7223,hollebeek:1999,sol00:4415,scholes:2000,hol01:3940,hol01:3945,ho03:6433,luo14:3099,MM.mbno:2015,unk17:1923,MM.n2o:2020,MM.h2co:2020,MM.rkhs:2020}\\

\noindent
To make an inference (prediction) of an unknown value $y_{*}$ at a
given value $\vec{r}_{*}$ one can use Bayes' theorem according to
which
\begin{equation}
P(y_{*} | \vec{y}) = \int_{\theta} P(y_{*} | \theta) P(\theta | \vec{y}) d \theta
\end{equation}
where $\theta$ is a vector containing the model parameters and
$\vec{y}$ are the known values (e.g. energies from {\it ab initio}
calculations or scattering cross sections).\cite{krems:2019} If the
model parameters $\theta$ are not fixed but random variables
themselves, a Bayesian NN is obtained. More specifically, if Gaussian
distributions are assumed for the parameters and in the limes of an
infinite number of neurons, the Bayesian NN has been shown to map onto
a GP.\cite{krems:2019}\\

\noindent
This formalism has been recently applied to reactive scattering for
the H+H$_2$$\rightarrow$H$_2$+H and the OH+H$_2$$\rightarrow$H$_2$O+H
reactions. It was reported that for the H+H$_2$$\rightarrow$H$_2$+H
reaction with as few as 37 points (taken from a total of 8701
reference energies from which a global PES had been fitted previously)
accurate reaction probabilities can be
obtained.\cite{krems.bayes:2019} Similarly, for the more challenging
OH+H$_2$$\rightarrow$H$_2$O+H reaction 290 points from a total of
$\sim 17000$ {\it ab initio} energies were sufficient to cover the
entire 6-dimensional PES to obtain accurate reaction probabilities as
a function of the translational energy of the
reactants.\cite{krems.bayes:2019}\\

\noindent
It is worthwhile to mention that with increasing dimensionality
GP-based representations become computationally
expensive.\cite{panesi:2020} On the other hand, increasing the kernel
complexity, e.g. by using composite kernels, both the
performance\cite{krems:2018,krems:2020} and
accuracy\cite{krems.jctc:2020,krems.jcp:2020} of this approach have
been found to improve.\\

\section{Reaction Rates and Pathways}
Reaction rates can be determined from classical or quantum dynamics
simulations if suitable PESs are available that allow bond formation
and bond breaking. Ideally, such PESs are {\it global}, i.e. they
allow - starting from a reactant structure - to form all chemically
meaningful and energetically accessible product states. In practice,
generating such global PESs is extremely
challenging\cite{bowman:2011,guo:2020} or even impossible due to the
large number of reaction pathways. The global nature of the PES is
particularly important in high-energy processes such as hypersonics or
in combustion. While for
hypersonics\cite{candler:1995,sarma:2000,cummings:2003,walpot:2012,leyva:2017,dsmc:2017,MM.reactions:2020}
the relevant species are often atoms and diatomics, this is not the
case for combustion\cite{klippenstein:2017} or for atmospheric and
astrophysically
relevant\cite{vereecken2018perspective,wakelam:2010,wakelam:2012,balucani:2009}
processes for which the species involved can be larger and the number
of possible product channels therefore increases considerably.\\

\noindent
To illustrate the problem for following a chemical reaction of an
atmospherically relevant molecule isomerization and decomposition
pathways for acetaldehyde (AA) are considered. These processes are
relevant for atmospheric chemistry because it has been proposed that
formic acid (FA) can be generated via oxidation by the hydroxyl
radical\cite{shaw2018photo,dasilva:2014} following
photo-tautomerization of AA to its enol form
VA.\cite{archibald:2007,kable:2012,osborn:2012} Pathways in addition
to the conventional route (photochemical oxidation of biogenic and
anthropogenic volatile organic compounds (VOCs)) for formation of FA
are required to account for the global budget of formic
acid.\cite{millet:2015}\\

\noindent
To characterize the isomerization between AA and VA under conditions
relevant to the atmosphere a NN-based, reactive PES was
constructed\cite{mm.atmos:2020} based on PhysNet.\cite{unk19:3678}
Such a PES is required to run statistically significant numbers of
trajectories based on QCT simulations\cite{MM.armd:2014,MM.h2so4:2014}
because ab initio MD simulations are computationally too
expensive. The excitation energy in the simulations was 93.6 kcal/mol
which compares with energies of 86.6 kcal/mol to 95.3 kcal/mol for
actinic photons. At this excitation energy no isomerization reaction
from AA to VA was observed but decomposition into CH$_4$+CO, and
H$_2$+H$_2$C$_2$O occurred. It was found that for an accurate
representation of all states involved (AA, VA, CH$_4$+CO, and
H$_2$+H$_2$C$_2$O) and for stable $NVE$ simulations more than $4
\times 10^5$ reference energies at the MP2 level of theory with an
aug-cc-pVTZ basis set were required.\cite{mm.atmos:2020} For energies
up to 93.6 kcal/mol (isomerization barrier between AA and VA at 68
kcal/mol and excitation energy by actinic photons) the MAE and RMSE
are 0.0071~kcal/mol and 0.0145~kcal/mol, respectively. In order to
validate that the NN-PES does allow isomerization, higher excitation
energies up to 127.6 kcal/mol were used. The global PES has then a MAE
and an RMSE of 0.0132~kcal/mol and 0.0307~kcal/mol, respectively.\\

\noindent
These simulations found that for excitation energies of $\sim 95$
kcal/mol - corresponding to actinic photons - not a single
isomerisation occurred on the 500 ns time scale. Hence, formation of
FA following electronic excitation with actinic photons of AA and
subsequent ground state relaxation and isomerisation to VA (and/or
further chemical processing) appear unlikely to occur, in contrast to
the interpretation of the experiments.\cite{shaw2018photo} Rather,
after photoexcitation of AA the system either decomposes or relaxes
through internal vibrational energy distribution.\\

\noindent
Another ML-based approach that was recently followed to construct
reactive PESs and use them in dynamics simulations is based on
permutationally invariant polynomials (PIP)\cite{bra09:577,qu18:151}
combined with a neural network (PIP-NN).\cite{guo:2016} For
conventional PIP, the expansion coefficients in the polynomials
(usually in Morse-variables) are fitted using a linear least squares
algorithm whereas for PIP-NN the coefficients are trained by a
NN. PIP-NN has been applied to both, gas-phase and surface reactions.\\

\noindent
In the gas phase, PIP-NN has been applied\cite{guo:2014} to reactions
such as HO$+$CO$\rightarrow$H$+$CO$_2$ which is relevant in the
atmosphere and in combustion.\cite{francisco:2010} A total of $\sim
75000$ was used to represent this channel. The PES was used in QCT and
quantum dynamics simulations to determine total reaction
probabilities, thermal rates, differential cross sections, and product
state vibrational and rotational distributions\cite{guo:2014} as well
as tunneling probabilities and survival fractions.\cite{guo.2:2014}
Comparison between a pure NN, a pure PIP, and the PIP-NN approaches
demonstrates that despite the rather small differences in the fitting
quality certain observables, such as product state distributions or
differential cross sections can sensitively depend on the shape of the
PES.\\

\noindent
PIP-NN has also been used for reactive scattering involving metal
surfaces. Systems investigated include H$_2$/Ag(111),\cite{jiang:2014}
H$_2$/Co(0001),\cite{jiang.2:2015} H$_2$O/Ni(111),\cite{jiang:2015}
and CO$_2$/Ni(100).\cite{jiang:2016} The number of reference points in
these applications ranged from several 1000 to $\sim 25000$. For the
dissociative chemisorption of water on rigid Ni(111), QCT simulations
using a nine-dimensional PIP-NN PES fitted to energies from density
functional theory it was found that the reactivity depends on the
impact sites and the incident angle of the water
molecule.\cite{jiang:2015} Furthermore, analysis of the simulations
demonstrated that both, the barrier height and the topography of the
PES influence the reaction rate as does the translational energy both,
parallel and perpendicular to the surface.\\

\noindent
More recently, PIP-NN has also been extended to larger systems,
including the F+ CH$_3$OH$\rightarrow$ HF$+$CH$_3$O
reaction\cite{neumark:2017} or the investigation of the
Cl$+$CH$_3$OH$\rightarrow$HCl$+$ CH$_3$O/CH$_2$OH
reaction.\cite{guo.2:2020} These examples illustrate that it is
possible to follow reactions with multiple product states in a
realistic fashion. For instance, the HCl vibrational and rotational
product distributions and the product translational energy
distributions compare well with experiment.\\

\begin{figure}
\includegraphics[scale=0.6]{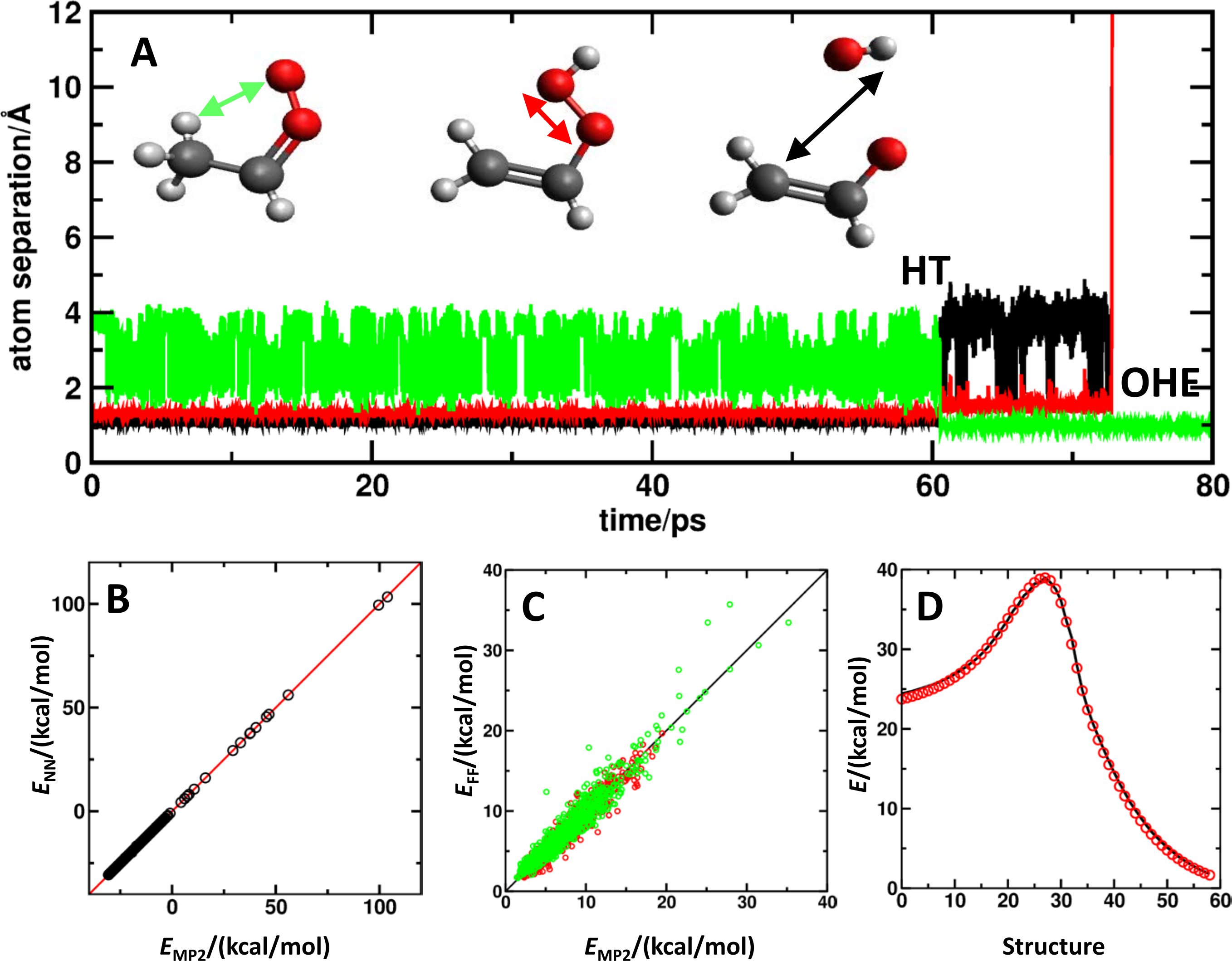}
\caption{Stepwise reactions for the Criegee intermediate. Panel A:
  Time dependence of important atom separations, the CH (black), OO
  (red), and OH (green), for the H-transfer (HT) and OH-elimination
  (OHE) reaction. Panel B: PhysNet\cite{unk19:3678} model for $\sim
  10^5$ MP2 reference energies together with the correlation between
  them (red). Panel C: MS-ARMD\cite{MM.armd:2014} model for $\sim
  100³$ reference structures for reactant (red) and product (green)
  structures. The RMSD is $\sim 1$ kcal/mol and the $1:1$ correlation
  is the black line. Note the different energy scales in panels B and
  C. Panel D: minimum energy path for H-transfer between reference
  (red circles) and MS-ARMD fit (black). Energy-dependent rates from
  both models for the energy (PhysNet and MS-ARMD) are in good
  agreement with experiment.\cite{lester.2:2016}}
\label{fig:criegee}
\end{figure}

\noindent
Complementary to reactions with multiple reaction products, multiple
step reactions with one or several intermediates between reactant and
product pose another challenge. A recent application concerned the
thermal activation of methane by MgO$^+$ for which experimental rates
were determined between 300 K and 600 K.\cite{MM.mgo:2020} Another
example is the unimolecular decomposition of the CH$_3$COOH Criegee
intermediate.\cite{lester:2016} Recent simulations based on an MS-ARMD
and NN-trained full-dimensional energy surface involving the reactant,
H-transferred intermediate, and OH-elimination product (see Figure
\ref{fig:criegee} demonstrate that stepwise reactions can also be
followed by such techniques. Figure \ref{fig:criegee} demonstrates
that empirical FFs can be fit with an accuracy of $\sim 1$ kcal/mol
(``chemical accuracy'') whereas using PhysNet to train the same data
reaches an accuracy of 0.02 kcal/mol. The preliminary rates from MD
simulations are consistent with experimental values.\\

\noindent
It should be emphasised that despite the undoubted accuracy of the
reactive PESs discussed above and their utility for interpreting
experiments and further elucidate the reaction dynamics of the
systems, they are not truly ``global'' PESs. Typically, the research
decides what channels are considered or of interest and the
corresponding breakup channels are included in the ML-based
construction of the interpolant. Thus, for systems with more degrees
of freedom and a larger number of product channels it is possible that
important states are omitted despite the powerful methods available
for (re)constructing the underlying PESs. If omitted, such channels
will fundamentally affect the resulting reaction network and the
realism with which one is able to map it to a computational model and
provide a meaningful complement to experiment.\\

\section{ML Applications to Reactive Biological Systems}

\subsection{Reactive Molecular Dynamics for Ligand Binding to Proteins}
Proteins are too large and the time scales of bond-breaking and
bond-forming reactions too long to be amenable to full {\it ab initio}
MD simulations. Hence, mixed QM/MM simulations which decompose the
system into a (small) reactive subsystem treated with quantum
mechanical methods and an environment described at the level of an
empirical force field are one of the methods of
choice.\cite{senn:2009,qiang.jcp:2016} Alternatively, methods such as
the empirical valence bond (EVB)
theory,\cite{warshel:1980,warshel:2003} or multi state adiabatic
reactive MD\cite{MM.armd:2014} have been developed and applied. More
recently, ML-based energy functions such as reproducing kernels
(RKHSs) have been used to follow bond-breaking and bond-formation in
biological systems. One example is nitric oxide binding to myoglobin
(Mb).\cite{MM.mbno:2015} For this, a 3-dimensional RKHS PES was fitted
to reference density functional theory calculations for radial and
angular degrees of freedom of the NO ligand with respect to the heme
unit and the iron out-of-plane motion with respect to the heme
plane. All remaining degrees of freedom of the solvated protein-ligand
system were treated with an empirical energy function.\\

\noindent
Extensive reactive MD simulations with such a mixed ML/empirical
energy function provided the first structural interpretation of the
metastable states in Mb-NO.\cite{MM.mbno:2016} Consistent with recent
optical and X-ray absorption experiments, which are unable to directly
relate spectroscopic response with the underlying structure, these
simulations found two processes: one on the 10 ps and another one on
the 100 ps time scale. They correspond to rebinding of the ligand to
Histidine64 in an ``open'' and a ``closed'' conformation. A mixed and
reactive ML/empirical energy function paired with the accuracy of RKHS
to represent the reference points is required to carry out the
necessary statistical sampling and reach the time scale of the
processes which is not possible with QM/MM techniques.\\

\subsection{Computer-Guided Enzyme Design}
Computer-based methods have also been used to modify or design amino
acid sequences that catalyze organic reactions.\cite{houk:2013}
Although a comprehensive and generally accepted ML-based technology to
do this is not available as of now, the recent success of
ML\cite{hassabis:2020} in winning the
CASP14\cite{deepmind.casp14:2020} clearly foreshadows that machine
learning will play an eminent role in design of enzymatically active
amino acid sequences. One of the deciding factors was the deeper
search capabilities of a CNN together with the choice of objective
function to be optimized which included distances between C$_{\beta}$
atoms, backbone torsion angles, and the prevention of steric
clashes.\cite{hassabis:2020} Thus, a combination of ``trainability''
through availability of curated and sufficiently complete reference
data, judicious choice of target information with respect to which
optimization can be carried out, and architecture of the underlying NN
are drivers for successful ML missions.\\

\noindent
Directed evolution has been used experimentally to (re)design enzyme
function.\cite{kern:2020} Starting from an originally designed Kemp
eliminase\cite{mayo:2012} the efficiency of the protein was assessed
after 7 and 17 rounds of evolution and found to have increased by more
than a 9 orders of magnitude. Analysis of nuclear magnetic resonance
(NMR) measurements indicated that the key difference between the
original and the evolved enzyme are the dynamic fluctuations of the
catalytic amino acids that increase the probability to occupy
catalytically proficient conformation and reduce the number of overall
possible conformations.\cite{kern:2020} Given the inherent
similarities of evolutionary strategies and neural networks it is
expected that ML-based technologies will provide further scope to
apply computer-based techniques for optimization and even reshaping
protein sequences for particular reactions. Solving such problems is
akin to the quest followed in materials design which aims at using ML
to develop materials with given properties (such as electrical
conductivity, melting point, or hardness).\\

\section{Machine Learning in the Context of Experiments}
Relevant experimental observables in the context of chemical
reactivity can include characteristics as diverse as the prediction of
reaction
probabilities.\cite{krems:2019,bowman.reaction:2019,bowman.reaction:2020},
differential cross sections or product state
distributions,\cite{MM.cross:2019,MM.reactions:2020} the prediction of
reaction outcomes (given specific input
compounds),\cite{aspuru:2016,rodrigues:2019} finding optimal reaction
conditions,\cite{zare:2017,jensen:2018} or predicting and identifying
fragmentation patterns from mass
spectroscopy.\cite{wishart:2016,bocker:2020} For prediction of reaction
outcomes it is worthwhile to mention that some of the efforts go back
at least 50 years with initial efforts to use computer-aided
strategies for organic synthesis
(CAOS).\cite{vleduts:1963,corey:1967,corey:1969,hendrickson:1971,gelernter:1973,gelernter:1977,jorgensen:1980}
Similarly, using ML-based techniques (referred to as ``AI'' at the
time) for analysis of mass spectrometric data started also in the
mid-1960s with
DENDRAL.\cite{lederberg:1965,lederberg:1993,djerassi:1969}\\

\subsection{Organic Reactions}
The field of ``retrosynthesis'' started around 1967.\cite{corey:1967}
Since then, much progress has been made. Initially, rule-based expert
systems such as CAMEO\cite{jorgensen:1980,jorgensen:1990} or
EROS\cite{gasteiger:1987} have attempted to predict reaction
outcomes. It was found that such approaches do not scale well and are
not easily generalizable. Later attempts were based on machine
learning approaches from training of labeled
reactions.\cite{kayala:2012} The ReactionPredictor uses a feature
vector consisting of physicochemical and topological features,
including molecular weight, formal and partial charges of the atoms,
information about atom sizes, together with information similar to
molecular fingerprints. For training the network, 1516 features were
retained. The learning was done within an artificial NN with sigmoidal
activation functions and one hidden layer, i.e. a shallow NN. For
organic textbook reactions as the training and the validation set
$\sim 96$ \% accuracy was reported.\cite{kayala:2012} Using a
fingerprint-based NN an accuracy of 80\% of selected textbook
reactions was found.\cite{aspuru:2016} The main limitation in this
prediction exercise was due to the limitations of the SMARTS
transformation to describe the mechanism of the reaction type
completely. Due to the flexibility in the descriptor it is possible to
further expand this algorithm to account for the reaction
conditions.\cite{aspuru:2016} Similarly, deep reinforcement learning
has been applied to optimize chemical reactions.\cite{zare:2017} For 4
different reactions it was shown that with the product yield as the
objective to maximize the Deep Reaction Optimizer (DRO) found the
optimal conditions within 40 steps, with the total time of 30 min
required to optimize a reaction in a microdroplet. Also, optimizing
reaction conditions for on one type of reaction and testing on a
different reaction (here the Pomeranz-Fritsch synthesis of
isoquinoline and the Friedl\"ander synthesis of substituted quinoline,
respectively) reached a higher yield with fewer optimization cycles.\\

\noindent
Despite these achievements, there are still limitations in using ML
methods to predict the outcome of diverse organic
reaction.\cite{gambin:2017} As an example of the intrinsic
difficulties faced one can consider the issue of ``learning'' in a
conventional ML context. Every ML model learns from a finite number of
training data, is tested on further independent data and then can be
used to predict on unknown data.\cite{unke:2020} Typically, the larger
the size of the training data, the better the model. There are on the
order of $10^7$ reactions with more than $\sim 10^4$ different
reaction types.\cite{grzybowski:2016} This leaves only $10^3$ samples
for every reaction which typically does not include different
solvents, reaction conditions, substitutions and other determinants
that drive a chemical reaction. Hence, the statistics for ``learning''
still needs to be improved.\\

\noindent
It will be of interest to see whether computer-assisted techniques can
contribute to alleviate such problems. Given the unparalleled increase
in computer efficiency, larger numbers of reactant, product and
transition state structures can be quantitatively evaluated
routinely. One example are the very large data sets employed to train
NN for molecular energy functions. The ANI-1 data set contains $2
\times 10^7$ structures of organic molecules.\cite{smi17:3192} For a
summary of existing databases, see Ref.\cite{huang:2020} With such
approaches it may be possible to more broadly assess structural,
substitutional and electronic effects on chemical reactivity that
undoubtedly are relevant for reaction outcomes. In addition, the
effects of solvent need to be included. There has been recent progress
in developing ML-based models for hydration and solvation free
energies on
compounds.\cite{jung:2019,keith:2019,bereau:2020,michel:2020} This,
together with the advances in electronic structure theory may provide
an avenue for further improvements of the models.\\

\noindent
Recently, a modular robotic system for organic synthesis consisting of
a Chemputer, a Chempiler and a scripting language (ChASM) was combined
to drive four modules consisting of a reaction flask, a filtration
station, a liquid-liquid separation module and a solvent evaporation
module.\cite{cronin:2019} This system was used to automate the
synthesis of compounds such as diphenhydramine hydrochloride,
rufinamide, or sildenafil without human
intervention.\cite{cronin:2019} Besides the attractive prospect to
automate standard chemical procedures with the opportunity to discover
new synthetic routes, such procedures also enhance the reproducibility
of synthetic procedures. An alternative approach of a robotics-based
platform driven by software uses From 12.5 million published
single-step reactions which were translated into a total of 163,723
rules.\cite{coley:2019} With this input a forward NN was trained to
predict what rules are most likely applicable for the synthesis of a
particular target molecule. The testing of the platform was done on 15
reactions of different complexity. It was concluded that such robotic
platforms coupled with curated data and powerful ML algorithms can
relieve scientists from routine tasks so that they can rather focus on
the more creative steps that lead to new ideas.\cite{coley:2019}\\

\noindent
Similar advances have been reported in the area of materials sciences
by introducing the concept of autonomous
experimentation\cite{roch:2018,aspuru:2020} based on
Phoenics\cite{aspuru.2:2018} and its successor
Gryffin.\cite{hase:2020} Phoenics uses Bayesian neural networks to
create a kernel-based surrogate model for the efficient optimization
of chemical and material properties. Like other Bayesian optimization
approaches, Phoenics balances the exploration and exploitation of
parameter space to achieve sample-efficient experimental
campaigns. Hence, such an approach also falls within the broader class
of ``model optimization problems'' that are also used for reactive
networks.\cite{sahinidis:2019}\\

\noindent
For the design of novel, functional materials with specific,
predefined properties, which - by definition - involve chemical
reactions and transformations, the problem at hand is further
exacerbated by the fact that the parameter space contains continuous
(e.g. $T$ or flow rates) and discrete (categorical) variables, such as
the type of solvent, chemical substitutions or the catalysts
used. Gryffin\cite{hase:2020}, an extension of the Phoenics framework,
allows to tackle such optimisation problems by taking advantage of
recent ML advances that enabled the continuous relaxation of discrete
variables\cite{maddison:2016,jang:2016}. This approach has been
applied to the autonomous optimization of a stereoselective
Suzuki-Miyaura coupling between a vinyl sulfonate and an arylboronic
acid to selectively generate the E-product isomer in high
yield.\cite{christensen:2020} Along similar lines, reaction yield
predictions were recently learned based on natural language
architectures using an encoder model and a regression layer. As an
example, the average $R^2$ for the Suzuki-Miyaura reactions was $\sim
0.8$, similar to a model only trained on the Buchwald-Hartwig
reactions. It was concluded that such models can perform equally well
on different types of reactions and are robust with respect to the
parameters and hyperparameters of the model.\cite{schwaller:2020}\\

\subsection{Mass Spectrometry}
Chemical structure determination using data from mass spectrometry was
one of the early applications of expert systems (``AI'') to problems
involving decomposition reactions.\cite{lederberg:1993} The earliest
ML-based program to do this was ``DENDRAL''.\cite{djerassi:1969} In a
later effort,\cite{gasteiger:1992} the insights gained from reaction
predictions using EROS\cite{gasteiger:1987} were used to develop a
computational framework (MAss Spectra SIMulatOr - MASSIMO) to predict
the mass spectrum given the structure of the compound. To put a
criticism of DENDRAL into context (``..it is sad to say that, in the
end, the DENDRAL project failed in its major objective of automatic
structure elucidation by mass spectral data..'')\cite{gasteiger:1992}
it should be noted that DENDRAL was aimed at the reverse task:
structure determination for given mass spectrometric data.\\

\noindent
More recently, NN-based techniques were developed to address the
problem of competitive fragmentation modeling for electron ionization
(CFM-EI).\cite{allen:2014,wishart:2016} Given a chemical structure the
model predicts an electron ionization (EI) mass spectrum
(MS). Contrary to another approach available,
CSI:FingerID,\cite{duhrkop:2015} competitive fragment modeling is
applicable to both, ions generated from electron ionization as well as
electrospray ionization (ESI). CFM-ESI uses a probabilistic model
based on systematic removal of all bond connections, every pair of
bonds in rings, and considering all hydrogen rearrangements within the
resulting fragments. The chemical features required for training the
NN include properties such as broken bond types (single, double, and
others), neighboring bond types, functional group
features,\cite{feunang:2016} and others.\cite{wishart:2016} The data
set for training, testing and validation contained $\sim 20000$
molecules. The performance of this model was 77 \% when querying
against the measured reference spectra and 43 \% against the NIST
database.\\

\noindent
Compound structure identification (CSI) in predicting fingerprints and
identifying metabolites (FingerID), referred to as
CSI:FingerID,\cite{duhrkop:2015} uses molecular fragmentation trees
with molecular fingerprint prediction based on multiple kernel
learning.\cite{heinonen:2012,shen:2014} Here, training was carried out
on $\sim 6200$ compounds. With the full training set the correct
identification rate is $\sim 30$ \%.\cite{duhrkop:2015} In a
comparative assay based on PubChem, the identification rate for
CSI:FingerID was $\sim 32$ \% compared with $\sim 12$ \% for
CFM-ID.\cite{duhrkop:2015}\\

\noindent
Most recently, analysis of high-resolution fragmentation mass spectra
was carried out based on ``class assignment and ontology prediction
using mass spectrometry'' (CANOPUS).\cite{bocker:2020} This workflow
employs a number of support vector machines (SVMs) to predict
fingerprints of the query compound which is the input to a deep neural
network to predict all possible compound classes consistent with the
query compound simultaneously. The SVMs are trained on experimental
reference mass spectrometric data. Conversely, the DNNs are trained on
millions of compound structures with molecular formulae as the feature
vectors together with the number of atoms of a given type, the mass,
and additional atom-based features as input to the
DNN.\cite{bocker:2020} The binary molecular fingerprint, determined
from CSI:FingerID,\cite{duhrkop:2015} and the molecular formula
features from a fragmentation tree\cite{bocker:2016} are used as input
to the DNN. The DNN is optimized using Adam.\cite{kingma2014adam} With
respect to performance as measured by the Matthews correlation
coefficient (${\rm MCC} = +1$ for perfect classification and ${\rm
  MCC} = -1$ for a completely wrong
classification)\cite{matthews:1975} the ranges are from 0.875 for
steroids to 0.972 for phosphocholines from the training set. For the
test set, the MCCs ranged from 0.60 to 0.74.\cite{bocker:2020}\\

\section{Machine Learning for Entire Reaction Networks}
Reaction networks are relevant in various branches of chemistry,
including but not limited to atmospheric reactions, combustion,
astrophysical and biological networks. Often such networks are sampled
at the level of a stochastic network\cite{gillespie:2007} by solving a
large number of coupled ordinary differential equations.\\

\noindent
More recently, it was attempted to directly propagate the nuclear
dynamics within an {\it ab initio} nanoreactor.\cite{martinez:2014}
Such an approach is still rooted in conventional {\it ab initio}
molecular dynamics simulations and limited to the level of theory
employed and the time scales accessible to such simulations. Very
recently, a NN-based model was presented to follow combustion
reactions in space and time.\cite{zhang:2020} These simulations used
the DeepMD NN architecture\cite{weinan:2018} to compute energies and
forces for methane combustion (starting 100 CH$_4$ and 200 O$_2$
molecules) at 3000 K and found 798 different chemical reactions, some
of which were as of now unknown.\cite{zhang:2020} The total simulation
time covered was in the nanoseconds and the accuracy of these
simulations is only limited by the electronic structure data the NN
was trained to.\\

\begin{figure}
\includegraphics[scale=0.95]{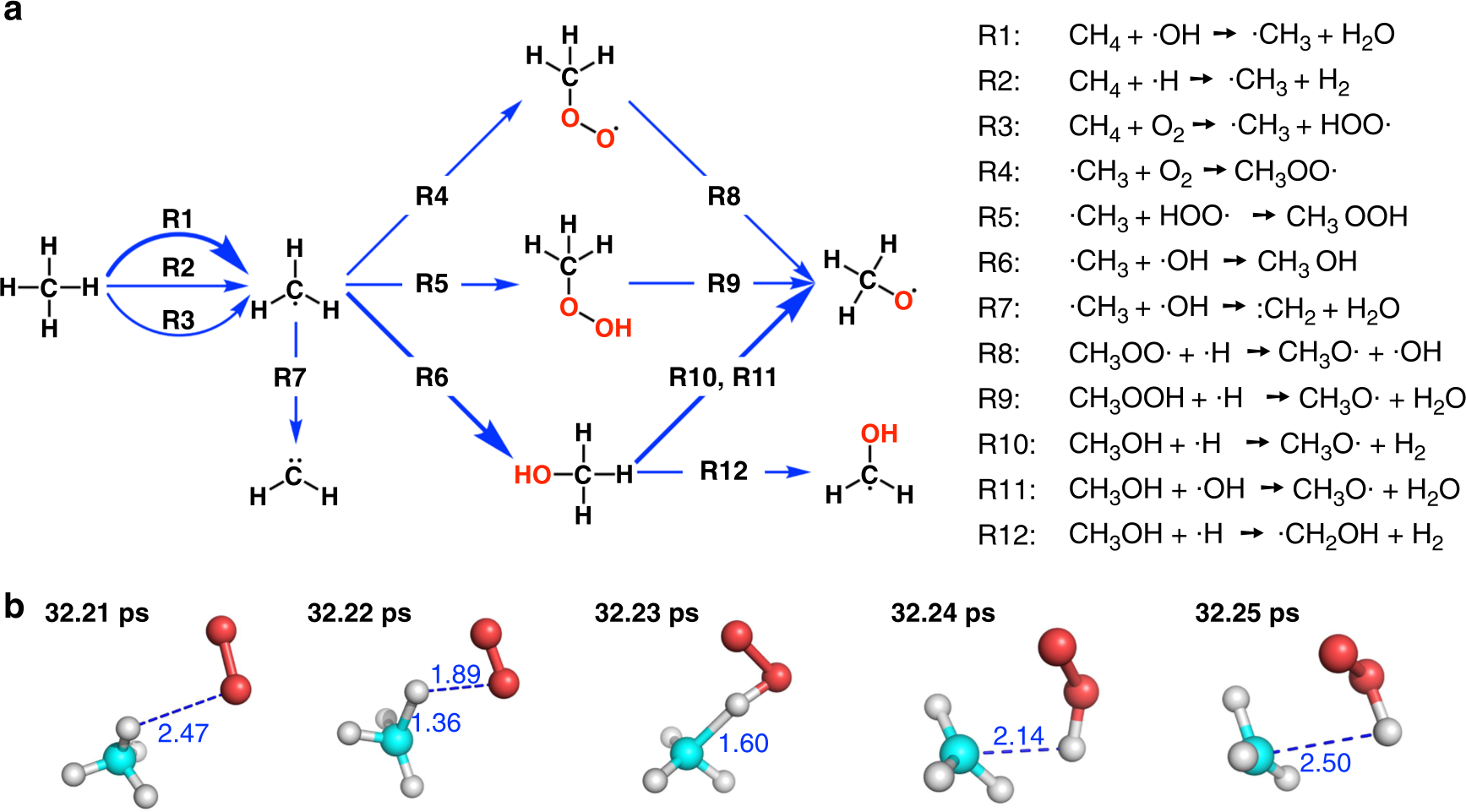}
\caption{The initial stage of methane combustion. Panel a: Primary
  reaction pathways (left) and reactions R1 to R1 (right) during the
  initial stage of the combustion. Panel b: A real-time trajectory
  showing the reaction progress over 40 fs for hydrogen abstraction
  from methane by O$_2$. Carbon, oxygen and hydrogen atoms are in
  cyan, red, and gray, respectively, with atom separations reported in
  \AA\/. Figure adapted with permission from Ref.\cite{zhang:2020}}
\label{fig:combustion}
\end{figure}

\begin{figure}
\includegraphics[scale=0.85]{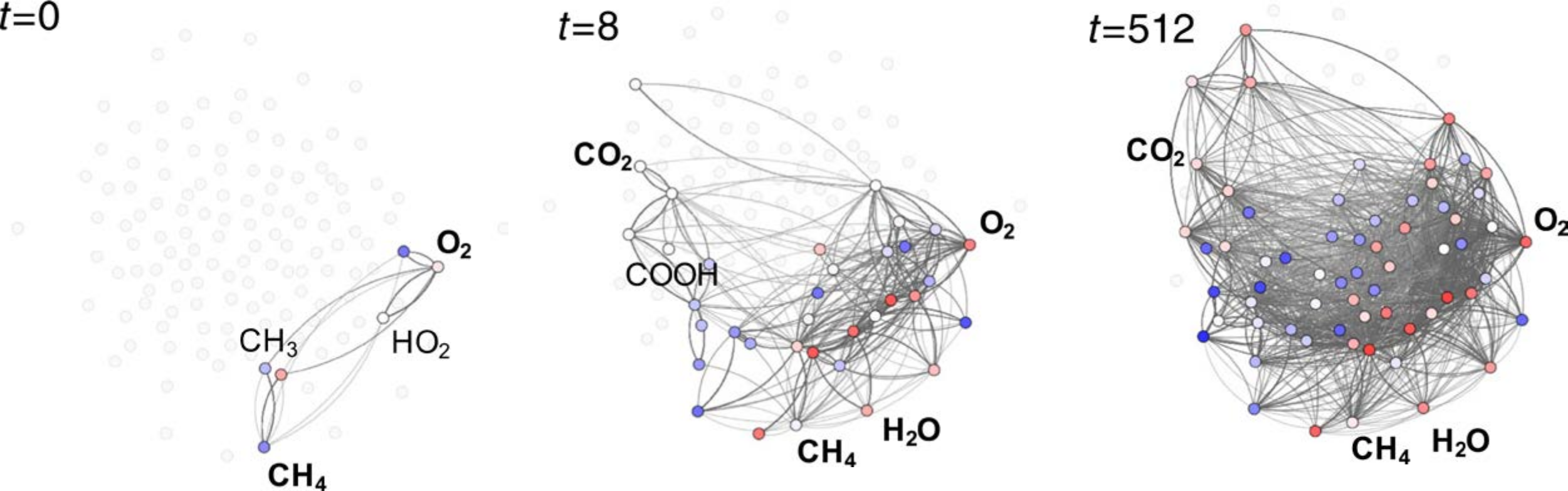}
\caption{Methane combustion: each frame shows the reduced reaction
  network extracted from a microkinetic simulation of methane
  combustion depending on abstract, arbitrary simulation time.
  Reactants and products (bold) and intermediates (regular font) are
  indicated next to the nodes which are colored according to their
  absolute atomization energies from low (red) to high (blue). Figure
  adapted from Ref.\cite{margraf:2020} with permission.}
\label{fig:methane}
\end{figure}

\noindent
In another recent attempt, methane combustion was simulated using an
ML-trained model on atomization energies\cite{margraf:2020} using
kernel ridge regression with a Smooth Overlap of Atomic Positions
(SOAP) representation.\cite{bartok:2013} A mean-field, qualitative
microkinetic simulation of a $50:50$ mixture of CH$_4$ and O$_2$ using
only the reaction energies (trained to an accuracy of $\sim 0.1$ eV)
and the law of mass action was carried out. The resulting reduced
reaction networks as a function of abstract simulation time is
reported in Figure \ref{fig:methane}. Several notable species are
formed in this simulation, including methanol, formic acid, and
Criegee intermediates.\\

\noindent
Machine-learning investigations of entire chemical reaction networks
were recently undertaken~\cite{sahinidis:2019}. Using ``automated
learning of algebraic models for optimization''
(ALAMO)~\cite{miller:2014}, the ``Reaction Identification and
Parameter Estimation'' (RIPE) tool was developed and used to estimate
and identify kinetic rate parameters from a postulated superset of
reactions. RIPE was applied to combustion reactors to model catalyst
conversion, or alternative reaction mechanisms and stoichiometric
relationships. Chemical looping combustion has been developed to
isolate fuel from air in combustion reactions using an oxygen carrier
shuttled between two flow reactors. In this application the task
solved by the ML approach is to select the model to best describe the
input data from experiment. For the application to alternative
reaction mechanisms the technique is used to discern between a large
number of alternative mechanistic pathways. Such approaches fall under
the general heading of ``optimal model selection'' given a concrete,
preconceived reaction network. The RIPE tool is available from
www.idaes.org and can handle between $10^2$ and $10^4$ reactions.\\

\section{Future Developments}
In the following possible developments in the field of ML-applications
to chemical reactions are illustrated.\\

\noindent
For small systems (containing few atoms) one question is whether in
general the number of reference points for constructing global,
reactive PESs can be dramatically reduced for accurate representations
of intermolecular PESs when resorting to ML techniques. Previously, it
was assumed that typically of the order of 10 points per degree of
freedom is required for a good coverage of conformational
space. Hence, for a diatom+diatom system of the order of $10^6$
reference {\it ab initio} energies would be required. For global,
reactive PESs this number is likely to be even larger. In the context
of GP regression it has been argued that for an $N-$dimensional system
only $\sim 10N$ well-selected points are required.\cite{loeppky:2009}
The work on combined ML and Bayesian optimization
techniques\cite{krems:2019} indicates that this is indeed possible for
molecular systems, too. On the other hand, recent work on the
SH+H$\rightarrow$S($^3$P)+H$_2$ reaction\cite{guo:2017} estimated that
rather 500 points are required for faithful representation of the
global, reactive PES, contrary to the $\sim 30$ points that were found
to be sufficient for the H+H$_2$$\rightarrow$H$_2$+H
system\cite{krems:2019} despite the same dimensionality. Hence, the
number of points required may depend on the presence of
(permutational) symmetry and the chemical species involved and hence
the overall topology of such a PES.\\

\noindent
One of the challenges ahead in the field is to learn high-quality PESs
while minimizing the number of reference points required. It has been
recently demonstrated that for high-dimensional, non-reactive systems
a few hundred points are sufficient to accurately represent the
near-equilibrium PES using RKHS representation on energies and
forces.\cite{MM.rkhs:2020} For the two largest molecules
(CH$_3$CONH$_2$ and CH$_3$COCH$_3$) 2500 reference energies were found
to be sufficient to obtain a mean averaged error of 0.01 and 0.07
kcal/mol on 1000 test points. The harmonic frequencies determined from
such PESs are typically within 1 cm$^{-1}$ of a normal mode
calculation using conventional normal mode analysis from quantum
chemical calculations at the same level of theory with a maximum
deviation smaller than 10 cm$^{-1}$.\\

\noindent
Another relevant question concerns the probing and specific
improvement of high quality PESs in view of experimental
observables. The question that arises in this context is which parts
of the PES are ``reliable'' and which parts can be further
improved. Chemical reactions by their very nature are sensitive to the
global shape of a PES whereas other observables such as harmonic
frequencies only probe the local shape and couplings between degrees
of freedom. The PES regions sampled for specific observables has,
e.g., been reported for the ${\rm N}(^4S) +{\rm O}_2(X^3\Sigma^-_g)
\leftrightarrow {\rm O}(^3P) + {\rm NO}(X^2\Pi)$
reaction.\cite{MM.no2:2020} Another study developed a Bayesian ML
approach to quantify uncertainties on PESs for the reactive O$_2$+O
system considering two different electronic states.\cite{panesi:2020}
This effort started from Bayesian-based sensitivity analysis of
computer models using GP.\cite{kennedy:2001} Sensitivity analysis of
PESs dates back at least 30 years\cite{rabitz:1992} where the problem
of inversion of ro-vibrational spectra for diatomic molecules has
already been formulated within a Tikhonov regularization framework.\\

\noindent
When using experimental observables to refine {\it ab initio}
calculated PESs for Ne-HF it has already been found that specific
observables are only sensitive to particular regions of the
PES.\cite{MM.morphing:1999,bowman:1991} Such ``PES morphing
approaches'' have been extensively applied to small molecular
systems.\cite{legon:1999,chandler:2000,jager:2000,bevan:2001,heaven:2001,tennyson:2001,barker:2001,hutson:2001,tennyson:2003,bowman:2003,xantheas:2004,belov:2004,bevan:2004,bevan:2008,spirko:2016,stolyarov:2016,spirko:2020}
There is also scope for extending this more broadly\cite{MM.pt:2019}
and in the context of machine learned PESs, as has been recently
indicated for acetylacetone.\cite{bowman:2020} Such approaches also
have the potential to more tightly integrate computation with
experiment and to develop computational models that learn from
experimental data.\\

\noindent
Besides the actual number of points it is also relevant to consider
the question what configurations of a system to use for the reference
calculations. The points should be pace in the most informative
regions, i.e. the regions the observables of interest are actually
sensitive to. This question as already been discussed in another
contribution to this special issue.\cite{manzhos:2020}\\

\noindent
For highly accurate, small molecule reaction dynamics based on
represented {\it ab initio} calculated PESs one remaining challenge is
the absence of a uniformly accurate and valid method to solve the
electronic Schr\"odinger equation. While for single-reference problems
coupled cluster (CC) techniques, such as CCSD(T), provide a valid
``gold standard'' such a generally applicable and robust technique is
missing in regions of the PES for which a single-reference electronic
wavefunction is not a good approximation. Multi-reference
configuration interaction (MRCI) methods are a viable alternative but
they are not yet of the same overall quality as CC-based techniques
and they can be challenging to apply. Also, computing entire PESs with
these methods can be non-trivial. It may be possible to apply mapping,
scaling and compound techniques to blend different methods across a
precomputed grid of interaction energies as has been done for example
for N$_3^-$.\cite{botsch:2013} One possibility are multi reference
coupled cluster (MRCC) methods although they remain computationally
challenging,\cite{evangelista:2018} in particular when global,
full-dimensional PESs are required. On the other hand, full
configuration interaction PESs have been computed for small,
few-electron systems\cite{MM.heh2:2019} and with increasing computer
speed such calculations and systems become tractable.\\

\noindent
Another area of future development concerns reactivity in
electronically excited states. For small systems, {\it ab initio}
calculations, ML-based representations of the PES and QCT dynamics
simulations similar to ground state problems have been used for rate
calculations for quite some
time.\cite{cas14:164319,alp17:2392,MM.cno:2018,MM.n2o:2020,MM.no2:2020}
However, for larger systems, the dynamics in the excited state is a
challenging problem in itself and ML-based techniques applied to this
(nonreactive) problem only start to appear.\cite{marquetand:2020} One
particular point of concern is the non-adiabatic dynamics and the
coupling matrix elements involved in the transition between
neighbouring electronic states.\cite{tully:2012,guo.2:2016}\\

\noindent
It is expected that combining currently available ML techniques for
fragmentation using mass spectrometry with advanced {\it ab initio}
calculations and data from existing data bases (see
Ref.\cite{huang:2020}) will further boost the quantitative side of
chemical structure determination from MS experiments. It is now
possible to determine optimized structures for entire chemical
libraries containing millions of
compounds\cite{reymond:2007,smi17:3192} at the density functional
theory level of theory. Such information is potentially useful for
better describing the reaction energetics of such decomposition and
fragmentation processes and the relative importance of each of the
channels.\\

\noindent
Visualization and virtual reality techniques for exploration of
chemical
reactivity\cite{wilson:1977,schulten:2001,martinez:2017,reiher:2014}
are other areas in which ML-based techniques may become
relevant.\cite{glowacki:2019} Such approaches are likely to also
become relevant in classroom
chemistry,\cite{aspuru:2018,rampino:2020,jesper:2019,glowacki:2019} A
personal note is related to first viewing an MD simulation of
myoglobin (using VMD) on a desktop screen as a postdoc working with
Prof. M. Karplus, It was fascinating to follow the dance of atoms when
nitric oxide was binding to and unbinding from the
heme-iron.\cite{MM.mbno:2002} Watching this reminded me immediately of
the notion ``..that everything that living things do can be understood
in terms of the jigglings and wigglings of atoms.''\cite{feynman:1965}
The impression of this shaped my personal view of chemical and
biological systems as inherently dynamical. Without capturing the
dynamics it was evident that a comprehensive understanding of chemical
and biological function\cite{mulholland:2018}, including enzymatic
activity\cite{kern:2020}, would not be possible. The combination of
electronic structure theory, molecular dynamics, machine learning and
virtual reality brings this one step closer and will also have
potentially far reaching, transformative effects on the way how we
will teach chemistry in the future.\cite{glowacki.2:2019}\\

\noindent
Machine learning techniques also have the potential to transform the
way in which the community regards the relationship between
experiment, simulation and theory. Together with automatization
(robotics), the combination of ML, experiment and simulation bears the
potential to develop integrated systems which optimize chemical
reaction systems with respect to a particular task (``loss
function''), such as maximizing yield, turnover, rate, or minimizing
use of problematic solvent. One of the fields which has seen many
advances is reaction planning for organic
synthesis\cite{rodrigues:2019} although examples for such system
optimization have been presented more than 20 years ago for gas-phase
reaction dynamics.\cite{rabitz.2:1992,rabitz:2011,rabitz:2020} in the
context of ``controlled chemistry''.\\

\noindent
In summary, ML-based approaches applied to chemical reactivity is a
rapidly expanding field. The challenges ahead concern the accurate,
quantitative and exhaustive determination of reaction outcomes, rates,
and (internal) state distributions. Coupled with robotic platforms,
reaction yields and reaction conditions can be optimized using ML and
Bayesian techniques. In the field of enzyme design, appreciable
improvements of turnover rates can be expected from coupling
experiments with ML-based approaches and for protein-ligand
interaction and recognition the recent advances made for protein
structure prediction will provide important insights. Finally,
exploring entire reaction networks has been possible very recently for
specific processes (e.g. methane oxidation). Together with improved,
high-quality reference data, exploration of chemical space for
reactions relevant to combustion, atmospheric sciences and
astrophysical chemistry will become viable.\\

\section*{Acknowledgment}
I am pleased to acknowledge the contribution from many of my students
and co- workers over the past decade to this research effort, in
particular Ms. Upadhyay for assistance with Figure
\ref{fig:criegee}. I also thank Profs. J. M. Bowman, H. Guo, R. Krems,
D. Wishart, N. Sahinidis, A. Aspuru-Guzik, and A. von Lilienfeld, and
Dr. M. Aldeghi for correspondence. This work has been financially
supported by the Swiss National Science Foundation (NCCR-MUST and
Grant No. 200021-7117810), the AFOSR, and the University of Basel.

\section*{Author Biography}
{\bf Markus Meuwly} studied Physics at the University of Basel and
completed his PhD in Physical Chemistry working with
Prof. J. P. Maier. After postdocs with Prof. J. Hutson (Durham) and
Prof. M. Karplus (Strasbourg and Harvard) as a Swiss National Science
Foundation Postdoctoral Scholar he started as a F\"orderprofessor at
the University of Basel in 2002 where he is Full Professor of Physical
and Computational Chemistry. He also holds a visiting professorship at
Brown University, Providence, RI. His scientific interests range from
accurate intermolecular interactions based on multipolar, kernel- and
neural network-based representations to applications of quantitative
molecular simulations for cold (interstellar) and hot (hypersonics)
environments and the investigation of the reactive dynamics and
spectroscopy in proteins and in the condensed phase. Several of the
tools are available in the CHARMM molecular simulation program.\\

\bibliography{refs}

\end{document}